\documentclass[superscriptaddress,amsmath,amssymb,twocolumn,nofootinbib]{revtex4-1}

\usepackage{amsmath, amsthm, amssymb, bm, braket}
\usepackage[dvips]{graphicx}
\usepackage{color}
\usepackage{wrapfig}
\usepackage{fancybox}
\usepackage{soul}

\DeclareMathAlphabet\bfcal{OMS}{cmsy}{b}{n}

\begin{document}

\title{Symmetry breaking of Gamow-Teller and magnetic-dipole transitions and its restoration in calcium isotopes}

\author{Tomohiro Oishi}
\email[E-mail: ]{tomohiro.oishi@yukawa.kyoto-u.ac.jp}
\affiliation{Yukawa Institute for Theoretical Physics, Kyoto University, Kyoto 606-8502, Japan.}
\affiliation{Department of Physics, Faculty of Science, University of Zagreb, Bijeni\v{c}ka cesta 32, HR-10000, Zagreb, Croatia.}

\author{Ante Ravli\'{c}}
\email[E-mail: ]{aravlic@phy.hr}
\affiliation{Department of Physics, Faculty of Science, University of Zagreb, Bijeni\v{c}ka cesta 32, HR-10000, Zagreb, Croatia.}

\author{Nils Paar}
\email[E-mail: ]{npaar@phy.hr}
\affiliation{Department of Physics, Faculty of Science, University of Zagreb, Bijeni\v{c}ka cesta 32, HR-10000, Zagreb, Croatia.}

\renewcommand{\figurename}{Figure}
\renewcommand{\tablename}{TABLE}
\newcommand{\crc}[1] {\hat{c}^{\dagger}_{#1}}
\newcommand{\anc}[1] {\hat{c}_{#1}}
\newcommand{\crb}[1] {\hat{a}^{\dagger}_{#1}}
\newcommand{\anb}[1] {\hat{a}_{#1}}
\newcommand{\CGC}[6] {\mathcal{C}^{(#1,#2){#3,#5}}_{#4,#6}} 
\newcommand{\JJJS}[6] {\left( \begin{array}{rrr} {#1}&{#3}&{#5}\\{#2}&{#4}&{#6} \end{array} \right)} 
\newcommand{\drsb}[2] {\bar{\psi}_{#1}(x_{#2})} 
\newcommand{\drsc}[2] {{\psi}^{\dagger}_{#1}(x_{#2})} 
\newcommand{\drsk}[2] {{\psi}_{#1}(x_{#2})} 
\newcommand{\bi}[1] {\ensuremath{ \boldsymbol{#1} }}
\newcommand{\oprt}[1] {\ensuremath{ \hat{\mathcal{#1}} }}
\newcommand{\abs}[1] {\ensuremath{ \left| #1 \right| }}
\newcommand{\slashed}[1] {\not\!{#1}} 
\def \beq{\begin{equation}}
\def \eeq{\end{equation}}
\def \beqa{\begin{eqnarray}}
\def \eeqa{\end{eqnarray}}
\def \bir{\bi{r}}
\def \ubir{\bar{\bi{r}}}
\def \bip{\bi{p}}
\def \ubip{\bar{\bi{r}}}
\def \adel{\ell} 
\newcommand \modific[2] {\textcolor{black}{{#2}}} 

\begin{abstract}
{Nuclear magnetic-dipole (M1) and Gamow-Teller (GT) transitions provide insight into the spin-isospin properties of atomic nuclei. By considering them as unified spin-isospin transitions, the M1/GT transition strengths and excitation energies are subject to isospin symmetry. The excitation properties associated to the M1/GT symmetry need to be clarified within consistent theoretical approach.
In this work, the relationship between the M1 and GT transitions in Ca isotopes is investigated in a unified framework based on the relativistic energy-density functional (REDF) with point-coupling interactions, using the relativistic quasi-particle random-phase approximation (RQRPA).
It is shown that the isovector-pseudovector (IV-PV) residual interaction affects both transitions, and 
the symmetry of M1 and giant-GT transitions is disrupted by this interaction in closed-shell nuclei.
In open-shell Ca isotopes, the proton-neutron pairing in the residual RQRPA interaction also plays a role in GT transitions.
Due to the interplay between these interactions, the M1/GT symmetry can be restored especially in the $^{42}$Ca nucleus, i.e., the giant-GT strength can become comparable to that of the M1 mode in terms of the unified spin-isospin transitions by adjusting the PN-pairing strength to reproduce the experimental low-lying GT-excitation energies.
The mirror symmetry of both M1 and GT transitions is also demonstrated for open-shell mirror partners, $^{42}$Ca and $^{42}$Ti.
Further improvements are required to achieve simultaneous reproduction of M1 and GT-transition energies in the REDF framework.
}
\end{abstract}

\maketitle

\section{Introduction} \label{Sec:intro}
Nuclear magnetic-dipole (M1) and Gamow-Teller (GT) modes have attracted interest in their isospin symmetry \cite{1968Ejiri,2013Ejiri}.
The GT operator for $A$-nucleon system formally reads 
$\sum_{k\in A} \hat{\tau}_{\pm}(k) \hat{\bi \sigma}(k)$, where $\hat{\tau}$ and $\hat{\sigma}$ indicate 
the isospin and spin, respectively.
For the spin-flip M1 transition, on the other side, its isovector (IV) operator is given as
$\sum_{k\in A} \hat{\tau}_{0}(k) \hat{\bi \sigma}(k)$ by neglecting physical factors.
Therefore, the GT and IV-spin M1 transitions can be interpreted as the unified spin-isospin transition, and 
a similarity based on the isospin symmetry is expected.
Theoretically, 
both the GT and M1 transitions are roughly understood by considering the spin-orbit (SO) splitting and relevant residual interactions \cite{1991Migli,1989Kamerd,2006Speth}: see also Refs. \cite{2009Vesely,2010Nest,2010Nest_2,2019Tselyaev,2020Speth,2020Kruzic,2020Oishi} for M1, and Refs. \cite{2003Vretenar,2004Paar,2004Ma,2012Niu,2012Sagawa_GT} for GT transitions.
In open-shell nuclei, the isoscalar and isovector pairing correlations have been shown as relevant \cite{2004Paar,2013Yoshida_GT,2014Bai,2019OP,2020Oishi,2020Minato,2021Vale,2021Ravlic_01,2021Ravlic_02}.


The GT transition is one of the main ingredients of beta radioactivity \cite{2007Suhonen,1992Osterfeld,2011Fujita_M1_GT}.
It plays an essential role in nucleosynthesis especially by determining the beta-decay lifetimes, which provide a key ingredient for understanding the r-process timescale \cite{1992Osterfeld,2011Fujita_M1_GT,KAJINO2019109}.
Furthermore, evaluation of the GT strength is important in order to predict the neutrino-induced reactions and electron capture in nuclei during the late stages of stellar evolution \cite{JANKA200738,2021Langanke}.
In Refs. \cite{2012Sagawa_GT,2014Bai}, the GT resonances are reproduced by improving the spin-isospin parameters in the Skyrme energy-density functional (EDF).
Recently in Ref. \cite{2020Gamba}, by utilizing the subtracted second Skyrme random-phase approximation (RPA), the GT strengths of $^{48}$Ca and $^{78}$Ni are obtained in better agreement with experiments than the other EDF calculations.
In addition, one remarkable advantage in studies on GT modes is the existence of the Ikeda-Fujii-Fujita sum rule \cite{1963Ikeda,1964Fujita}.
This rule has provided an essential constraint to validate studies of GT resonances and spin-isospin properties in nuclei \cite{2020Gamba,1982Bertsch,2002Jacek,2004Ma,2003Vretenar,2004Paar,2007Paar_rev,2012Niu,2012Sagawa_GT,2013Yoshida_GT,2014Bai}.

The M1 transition is the leading mode of magnetic transitions \cite{2007Suhonen,2010Heyde_M1_Rev,2008Pietralla_Rev,1985Rich,1990Rich}, and has been theoretically studied by utilizing 
the non-relativistic RPA \cite{2008Shimizu,2009Cao,2016Goriely,2009Vesely,2010Nest,2010Nest_2,2019Tselyaev,2020Speth},
relativistic RPA \cite{2020Kruzic,2020Oishi},
shell-model calculations \cite{2005Otsuka,2004Lang,2012Loens,2015Matsubara}, etc.
For further information, see the references in reviews \cite{2010Heyde_M1_Rev,2008Pietralla_Rev}.
In Refs. \cite{2004Lang,2012Loens}, the analogy with neutrino-nucleus scattering is discussed.
It also plays a role in the determination of neutron-capture rates, of significance for the r-process nucleosynthesis \cite{2004Goriely,2011Goriely,2012Loens}.
In Ref. \cite{2015Matsubara}, the minor (finite) quenching of the IS (IV) M1 strength is concluded.
There exists Kurath's sum rule for the M1 transitions in the case where the single-orbit approximation is applicable \cite{1963Kurath}.
In the non-relativistic RPA studies \cite{2009Vesely,2010Nest,2010Nest_2,2019Tselyaev,2020Speth}, the RPA-residual interaction essentially contributes in determining the M1 energy and strength, and its importance is also concluded in the relativistic RPA studies \cite{2020Kruzic,2020Oishi}.

For experimental measurement of GT transitions, two procedures have been utilized, namely, those from beta radioactivity and charge-exchange (CE) reactions.
The beta radioactivity may provide direct information on the GT strength $B_{\rm GT}$ \cite{2007Suhonen,1992Osterfeld}.
However, its experimental accessibility is limited by the energy release (Q value).
The CE reactions by strong-interaction probes, on the other side, enable one to measure the GT transitions toward higher energies.
The famous examples include the CE excitations by the $(p,n)$ as well as $(^{3}\text{He},t)$ reactions \cite{1985Ander,1997Wakasa,2006Ichimura,2007Fujita_GT_Exp,2011Fujita_M1_GT,2014Fujita}.
For M1 transitions, although its concept originates in the electro-magnetic interaction, 
those can be measured also with the common strong-interaction probes \cite{1989Crawley,2016Birkhan,2019Cosel,2011Fujita_M1_GT,2015Matsubara}.
Namely, assuming the isospin symmetry, this type of M1 transition can be interpreted as the isobaric-analog component in the same category of GT transitions.
Recent developments of experiments provide a rich amount of data to this aim.
However, the expected symmetry between these M1 and GT transitions has not been completely validated.
In addition, for improvement of theoretical models, the simultaneous reproduction of these M1/GT properties could be used as a standard benchmark.

In this work, we discuss the M1 and GT transitions within a common framework of the relativistic energy-density functional (REDF) theory \cite{1974Walecka,1977Boguta,1989Reinhard,2005Vret,2006Meng,2007Paar_rev,2011Niksic_rev,2016Ebran_PRC}. 
By utilizing the relativistic quasi-particle random-phase approximation (RQRPA) \cite{2003Paar,2004Paar,2005Niksic}, 
we discuss the relationship between M1 and GT transitions. The validity of current REDF parameterization is examined from the experimental M1 and GT data.

In this paper, we use the CGS-Gauss system of units. 
Also, all the systems discussed in the following sections are assumed as spherical.

\section{Formalism} \label{Sec:Form}
The relativistic mean-field calculation and RQRPA developed in Refs \cite{2007Paar_rev,2011Niksic_rev,2014Niksic,2020Kruzic,2020Oishi} are utilized in this work.
We briefly review the formalism and setting of parameters.
Our calculation is based on the point-coupling REDF determined from the Lagrangian density,
\beq
 \mathcal{L} = \mathcal{L}_{\rm PC} +\mathcal{L}_{\rm IV-PV},
\eeq
where $\mathcal{L}_{\rm PC}$ includes the isoscalar-scalar, isoscalar-vector, and isovector-vector coupling terms with point coupling (PC) interaction \cite{2008Niksic,2014Niksic}.
In this work, two parameterizations of density-dependent PC interactions are used, DD-PC1 \cite{2008Niksic} and DD-PCX \cite{2019Yuksel}.
In addition, for description of GT and M1 transitions, the isovector-pseudovector (IV-PV) Lagrangian density is included, where it contributes to the RQRPA residual interaction \cite{2005Vret, 2005Niksic, 1996Podo}.
That is, 
\beq
  \mathcal{L}_{\rm IV-PV} = -\hbar c\frac{\alpha_{\rm IV-PV}}{2}
  \left[ \bar{\psi} \gamma_5 \gamma_{\mu} \vec{\tau} \psi \right]
  \left[ \bar{\psi} \gamma_5 \gamma^{\mu} \vec{\tau} \psi \right]. \label{eq:AIVPV}
\eeq
Note that, since this IV-PV term leads to the parity-violating mean-field at the Hartree level, it does not contribute to the solution of natural-parity states, including the $0^+$ ground state.
In the non-relativistic limit, the IV-PV interaction derived from this Lagrangian density corresponds to the spin-isospin term in the Landau-Migdal interaction \cite{2020Oishi}.

\begin{table}[tb] \begin{center}
\caption{Interactions used in our RHB and RQRPA calculations. The label ph (pp) indicates the quasi-particle quasi-hole 
(quasi-particle quasi-particle) channel.}
  \catcode`? = \active \def?{\phantom{0}} 
  \begingroup \renewcommand{\arraystretch}{1.4}
\begin{tabular*}{\hsize} { @{\extracolsep{\fill}} ccccc} \hline \hline
&&M1 &&GT  \\  \hline 
  RHB ($0^+$) &ph &\multicolumn{3}{c}{DD-PC1/DD-PCX} \\
              &pp &\multicolumn{3}{c}{T1 pairing}  \\  \hline 
  RQRPA ($1^+$)      &ph &\multicolumn{3}{c}{DD-PC1/DD-PCX plus IV-PV} \\
                     &pp &T1 pairing  &&PN (T0) pairing  \\  \hline \hline
\end{tabular*}
  \endgroup
  \catcode`? = 12
\label{table:3172}
\end{center} \end{table}

For open-shell nuclei, pairing correlations in the particle-particle (pp) channel should be taken into account. 
For this purpose, we employ the same model used in Refs \cite{2005Vret,2008Niksic,2014Niksic,2020Kruzic}. 
Namely, for isospin-triplet (T1) pairs of proton-proton and neutron-neutron, a finite-range two-Gaussian potential is used.
That is,
\beqa
 V_{\rm pp,T1}(d) = && \sum_{i=1,2}  e^{-d^2 /\mu^2_i} \left( W_{i}+B_{i}\hat{P}^{\sigma} \right. \nonumber \\
 && \left. -H_{i}\hat{P}^{\tau} - M_{i}\hat{P}^{\sigma}\hat{P}^{\tau}  \right),
\eeqa
where $d=\abs{\bir_2-\bir_1}$. 
The operators $\hat{P}^{\sigma}$ and $\hat{P}^{\tau}$ indicate the exchanges of the spin and isospin, respectively.
Its parameters $\mu_{i}$, $W_{i}$, $B_{i}$, $H_{i}$ and $M_{i}$ are taken from the central part of Gogny-D1S force in Ref. \cite{1991Berger}.
Therefore, both in the M1 and GT cases, the ground state is obtained from the relativistic Hartree-Bogoliubov (RHB) method with the DD-PC1 and T1-pairing parameters in the particle-hole and particle-particle channels, respectively. 
This T1 pairing is also used in the RQRPA for M1 transitions.

For isospin-singlet (T0) proton-neutron pairing, on the other side, we assume the same potential used in the PN-RQRPA calculations in Refs \cite{2004Paar,2020Yuksel,2020Ravlic}, i.e., the similar two-Gaussian potential but only active in the $S_{12}=1$ channel.
That is,
\beq
V_{\rm pp,T0}(d) = f \left[ -G_0 \sum_{i=1,2} e^{-d^2 /\nu^2_{i}} g_i  \right] \Pi_{S_{12}=1,T_{12}=0},
\label{eq:PNPAIR}
\eeq
where $\Pi_{S_{12}=1,T_{12}=0}$ indicates the projection into the $(S_{12}=1,T_{12}=0)$ channel.
Here parameters $G_0$, $g_i$ and $\nu_i$ are the same as utilized in Ref. \cite{2020Ravlic} but with the strength parameter $f=0.81$ adjusted to reproduce the low-lying GT($-$) 
excitation energy of $^{42}$Ca, $0.611$ MeV \cite{2020Fujita}, which is shown in Sec. \ref{sec:openshell}.
We have summarized the interactions used for the ground-state (RHB) as well as the excited-state (RQRPA) calculations in Table \ref{table:3172}.

We use the isospin convention as $\hat{\tau}_0(k) \ket{k} = \tau_0 \ket{k}$ with $\tau_0=1~(-1)$ when the $k$th nucleon is neutron (proton).
The GT operator thus reads 
\beq
 \oprt{O}^{\rm GT(\pm)}_{\nu} = \sum_{k \in A} \hat{\tau}_{\pm}(k) \hat{s}_{\nu}(k), \label{eq:CIRYGE}
\eeq
where $\nu=\pm 1$ or $0$ (magnetic quantum number). 
The isospin operator $\hat{\tau}_{\pm}(k)$ changes the charge of $k$th nucleon in the GT($\pm$) transition.
\modific{2}{Notice that the spin operator $\hat{s}_{\nu}$ is used instead of $\hat{\sigma}_{\nu}=2\hat{s}_{\nu}$, and thus, the present definition is different by factor $1/2$ from the usual GT operator.}
On the other side, the isovector (IV) M1 operator reads
\beq
\oprt{O}^{\rm M1}_{\nu} = \mu_{\rm N} \sqrt{\frac{3}{4\pi}} \sum_{k \in A} \hat{\tau}_{0}(k) \left[  g_s \hat{s}_{\nu}(k)  +g_l \hat{l}_{\nu}(k)  \right], \label{eq:SPWUE}
\eeq
where $\mu_{\rm N}$ indicates the nuclear magneton, and the nuclear g factors for the IV M1 mode are $g_s=4.706$ and $g_l=1/2$ \cite{1963Kurath,70Eisenberg}.
However, with respect to the GT operator,
we neglect the orbital-M1 component, $\hat{\tau}_0 \hat{l}_{\nu}$, except when mentioned.
We also omit the factors $g_{s}$ and $\mu_{\rm N} \sqrt{3/4\pi}$ in the following sections.
This omission corresponds to the renormalization of $B_{\rm M1}$ as in Ref \cite{1997Fujita_M1_GT}.
\modific{1}{Namely, we employ the IV spin-M1 operator,
\beq
\oprt{O}^{\rm M1S}_{\nu} = \sum_{k \in A} \hat{\tau}_{0}(k) \hat{s}_{\nu}(k), \label{eq:CIRZPE}
\eeq
which is in correspondence to the GT operator.}
In this work the so-called quenching factors on M1 and GT operators are not 
considered for simplicity.

For the orbital-M1 component, $\hat{\tau}_0 \hat{l}_{\nu}$, its contribution to the IV M1 response is indeed finite, but it does not change the M1-excitation energies.
We describe its details in the Appendix.
In the main text, we neglect this ``inclusion'' into the unified spin-isospin modes, and consider their operators as in Eqs. (\ref{eq:CIRYGE}) and (\ref{eq:CIRZPE}).

The charge-conserving RQRPA is utilized for the IV-spin-M1 mode.
Using the QRPA ansatz, the M1-excited state $\ket{\omega}$ is formally given as 
\beq
 \oprt{H} \ket{\omega} =\hbar \omega \ket{\omega},~~~~\ket{\omega} =\oprt{Z}_{\rm M1}^{\dagger}(\omega) \ket{\Phi}, \label{eq:2654211}
\eeq
where $\ket{\Phi}$ is the RHB ground state and $\hbar \omega$ is the excitation energy.
The M1-excitation operator reads 
\beq
  \oprt{Z}_{\rm M1}^{\dagger}(\omega) = \sum_{\rho <  \sigma}
  \left\{   X_{\rho \sigma}(\omega) \oprt{Q}_{\sigma \rho}^{\dagger}
         -Y^*_{\rho \sigma}(\omega) \oprt{Q}_{\sigma \rho}
  \right\},
\eeq
with $\oprt{Q}_{\sigma \rho}=\left[\anb{\sigma} \otimes \anb{\rho} \right]^{(J,P)}$ coupled to the $J^P=1^+$ spin and parity.
Here $\anb{\sigma}$ is the quasi-particle operator with the label $\sigma$ for quantum numbers.
The labels $(\rho,\sigma)$ indicate proton-proton and neutron-neutron pairs.
\modific{3}{In the PN-RQRPA for GT($\pm$) modes \cite{2004Paar,2005Niksic}, on the other side, it reads
\beq
  \oprt{Z}_{\rm GT}^{\dagger}(\omega) = \sum_{\alpha, \beta}
  \left\{   X_{\alpha \beta}(\omega) \oprt{Q}_{\beta \alpha}^{\dagger}
         -Y^*_{\alpha \beta}(\omega) \oprt{Q}_{\beta \alpha}
  \right\},
\eeq
where labels $(\alpha,\beta)$ indicate proton-neutron $1^+$ pairs.}
Then, by solving the matrix form of the QRPA equation, excitation amplitudes are obtained:
\beq
  \left( \begin{array}{cc}  A  &B  \\ B^*  &A^* \end{array} \right)
  \left( \begin{array}{l}  X(\omega)  \\ Y^*(\omega) \end{array} \right)
  =
  \hbar \omega
  \left( \begin{array}{cc}  I  &0  \\ 0  &-I \end{array} \right)
  \left( \begin{array}{l}  X(\omega)  \\ Y^*(\omega) \end{array} \right), \label{eq:qrpa7131}
\eeq
where $A$ and $B$ are the well-known QRPA matrices \cite{80Ring,2003Paar,2005Niksic}. 
From the RQRPA solutions $\ket{\omega_{f}}$ with $E_{f}=\hbar \omega_{f}$, response functions can be obtained as
\beqa
R_{X}(E)
&=& \sum_{f} \delta(E-E_{f}) B_{X}(E_f) \nonumber \\
&=& \sum_{f} \delta(E-E_{f}) \sum_{\nu=\pm 1,0} \abs{\Braket{f \mid \oprt{O}^{X}_{\nu} \mid \Phi}}^2, \label{eq:response}
\eeqa
for $X={\rm M1S}$ and GT($\pm$).
For the purpose of visualization, these response functions are smeared with the normalized Cauchy-Lorentz profile, whose full width at half maximum (FWHM) is $1.0$ MeV in this work.
For magic nuclei, corresponding to the zero-pairing limit, the RQRPA reduces to the relativistic random-phase approximation (RRPA).
Similarly to Refs. \cite{2020Kruzic,2020Oishi}, our RHB and R(Q)RPA calculations are performed in the model space expanded in the harmonic oscillator (HO) basis up to 20 major shells.
The cutoff energy $120$ MeV for the configuration space in the QRPA is checked to be sufficient to have the convergence of results.

\begin{figure}[t] \begin{center}
\includegraphics[width = \hsize]{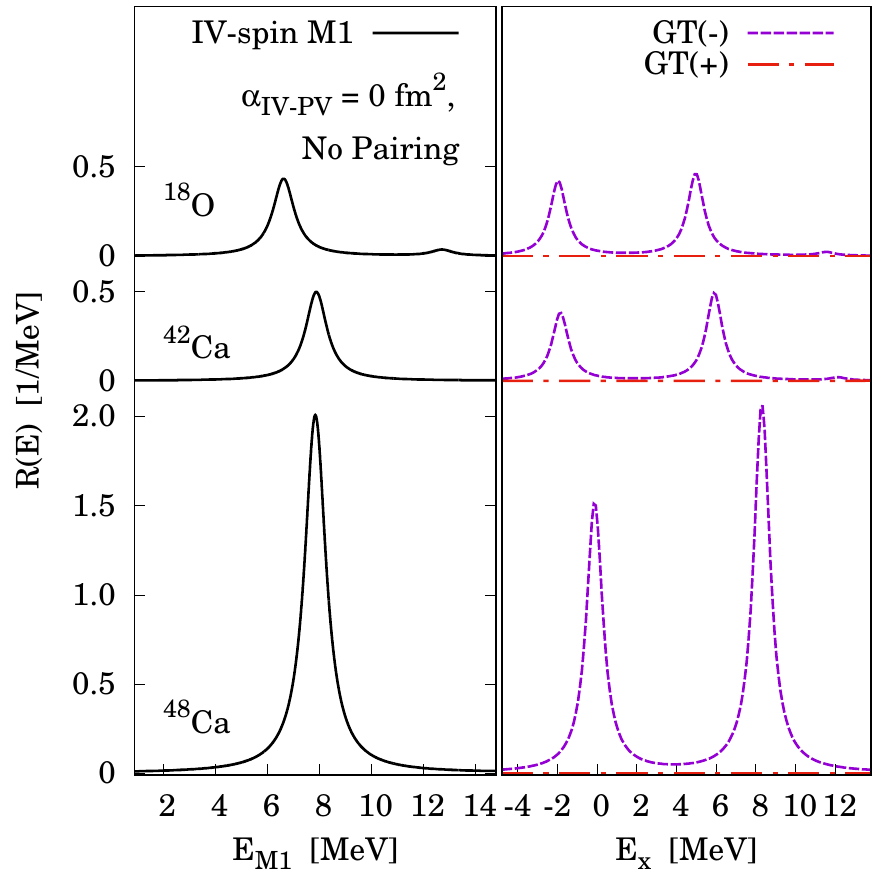}
\caption{The non-perturbed IV-spin-M1 and GT($\pm$) strength distributions obtained with the DD-PC1 interaction.
The GT($\pm$)-excitation energies $E_{\rm x}$ are presented with respect to the daughter nuclei \cite{2020Yuksel}.
} \label{fig:0258}
\end{center} \end{figure}

\section{Results} \label{Sec:Results}
\subsection{Non-perturbed response}
Before going to the full consideration of the RQRPA calculations, first we show the non-perturbed responses for the GT and M1 transition operators.
They are obtained by neglecting the residual RQRPA interaction, thus keeping only the response calculated with the DD-PC1 interaction.
The pairing (pp) correlations in the ground states are also neglected.

In Fig. \ref{fig:0258}, the non-perturbed response functions for GT and M1 modes obtained with the DD-PC1 interaction are displayed for $^{18}$O, $^{42}$Ca, and $^{48}$Ca.
Since their operators introduced in Eq. (\ref{eq:CIRYGE}) have the identical form of $\hat{\tau}\hat{s}$, their responses can be directly compared.
Note that, for GT($\pm$) modes, their excitation energies are given with respect to the ground states of daughter nuclei by using the converting method in Ref. \cite{2020Yuksel}.

First we focus on the isobaric-analog symmetry between the IV-spin-M1 and the giant, higher-energy GT($-$) transitions.
For all three nuclei shown in Fig. \ref{fig:0258}, the non-perturbed IV-spin-M1 strength is equal to the higher-energy GT($-$) strength: $R_{\rm M1} \cong R_{\rm GT(-)}(E_{\rm x})$ with $E_{\rm x}=5.4$, $6.6$, and $8.5$ MeV in $^{18}$O,
$^{42}$Ca, and $^{48}$Ca, respectively.
By analyzing the transition components, this equivalence can be explained from the identical set of quasi-particle transitions.
That is,
\beqa
&& \nu(1d_{5/2}) \longrightarrow \nu(1d_{3/2})~~{\rm in}~{\rm IV~M1},~{\rm and} \nonumber  \\
&& \nu(1d_{5/2}) \longrightarrow \pi(1d_{3/2})~~{\rm in}~{\rm GT(-)}, \nonumber
\eeqa
for $^{18}$O, where the symbol $\pi$ ($\nu$) indicates the proton (neutron) state. Similarly for $^{42,48}$Ca, 
\beqa
&& \nu(1f_{7/2}) \longrightarrow \nu(1f_{5/2})~~{\rm in}~{\rm IV~M1},~{\rm and} \nonumber  \\
&& \nu(1f_{7/2}) \longrightarrow \pi(1f_{5/2})~~{\rm in}~{\rm GT(-)}. \nonumber
\eeqa
Because the form of operators and the set of quantum labels are identical, 
it is natural to produce the same amount of transition strength in these two modes. 

On the GT($-$) side in Fig. \ref{fig:0258}, there is another, low-lying peak.
This low-lying GT response is attributable to the charge-exchange transition of $\nu(j=l+1/2) \to \pi(j=l+1/2)$ \cite{2004Paar}.
That is, for $^{18}$O,
\beq
 \nu(1d_{5/2}) \longrightarrow \pi(1d_{5/2})~~{\rm in}~{\rm GT(-)}, \nonumber
\eeq
whereas, for $^{42,48}$Ca,
\beq
 \nu(1f_{7/2}) \longrightarrow \pi(1f_{7/2})~~{\rm in}~{\rm GT(-)}. \nonumber
\eeq
Notice that its isobaric-analog transition of M1-type is nonphysical because the initial and final states are in this case identical.

In Fig. \ref{fig:0258} the non-perturbed low-lying GT($-$) responses show the negative excitation energies.
This implies that, by neglecting the residual RQRPA interactions, the calculated GT($-$) energies of daughter nuclei inevitably result lower than the experimental ground-state energies.
This drawback is remedied in the next section by including the residual interactions.

By comparing the non-perturbed M1 and GT($-$) energies in Fig. \ref{fig:0258}, one can observe that $E_{\rm M1}\cong \Delta E_{\rm GT(-)}$ for each nucleus, where $\Delta E_{\rm GT(-)}$ is the gap between the higher and lower GT($-$) energies.
This is trivial because $E_{\rm M1}$ ($\Delta E_{\rm GT(-)}$) corresponds to the SO-splitting gap of relevant orbits of neutrons (protons).
Since the relevant SO splitting is of the same $lj$ numbers, the $E_{\rm M1}$ and $\Delta E_{\rm GT(-)}$ values are also similar.

For these nuclei, the GT($+$), as well as proton-M1 transitions, 
are strongly suppressed by the Pauli-blocking effect.
By comparing the results for Ca isotopes, one can read that 
the non-perturbed M1 and GT responses are proportional to 
the number of valence neutrons around the $^{40}$Ca core, namely, 
$B_X(^{48}{\rm Ca}) \cong 4B_X(^{42}{\rm Ca})$ both for $X=$ M1 and GT.
This is simply because the $^{40}$Ca nucleus cannot be active for M1 neither GT transitions.

Note that, for the $^{18}$O and $^{42}$Ca nuclei, the two-valence-nucleon M1 sum rule \cite{2019OP} was consistently reproduced in the non-perturbed relativistic RPA \cite{2020Kruzic}.
For the $^{42}$Ca and $^{48}$Ca nuclei, in addition, the Kurath's M1 sum rule \cite{1963Kurath} was also reproduced:
its details were presented in Ref. \cite{2020Oishi}.
On the GT side, the Ikeda-Fujii-Fujita sum rule \cite{1963Ikeda, 1964Fujita} can be reproduced up to $90$-$95$\% with a small deficiency due to omitting the Dirac-sea states at the R(Q)RPA level \cite{2003Paar,2004Paar,2005Niksic}.

\begin{figure}[t] \begin{center}
  \includegraphics[width = \hsize]{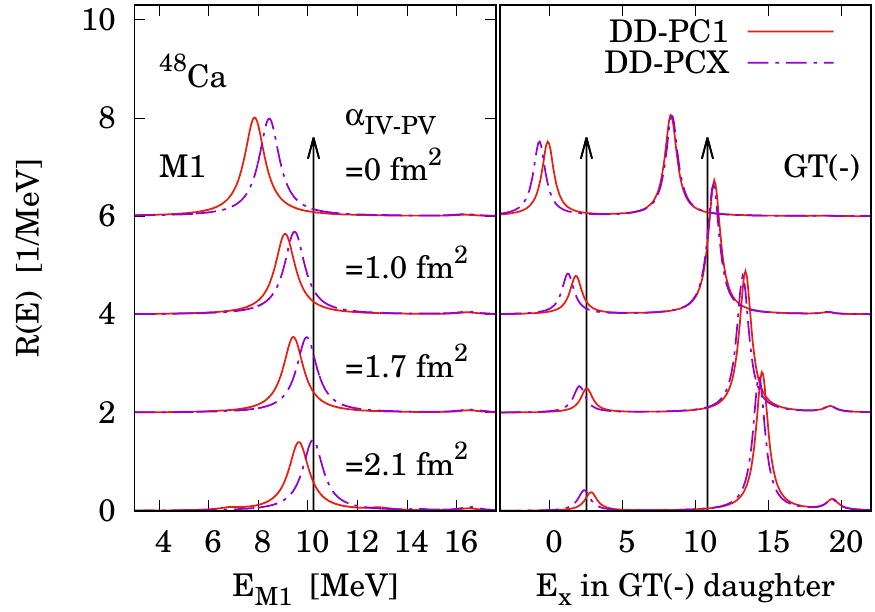}
  \includegraphics[width = \hsize]{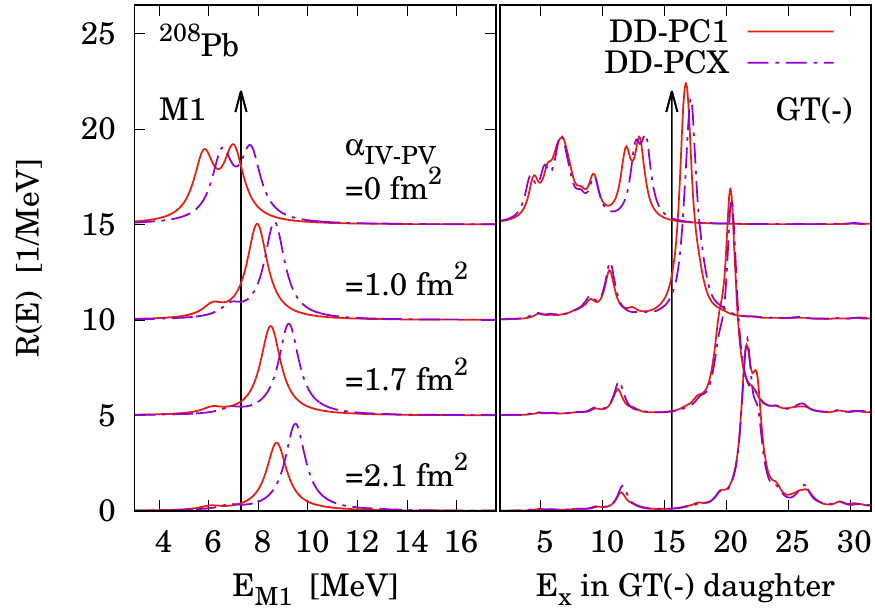}
  \caption{(Top panel) The IV-spin-M1 and GT($-$) strength distributions for $^{48}$Ca with the DD-PC1 and DD-PCX interactions supplemented with the IV-PV residual interaction in the RRPA.
The arrows indicate the experimental M1 and GT($-$) energies: $E_{\rm M1}=10.23$ MeV \cite{2011Tompkins}; $E_{\rm GT}=2.517$ MeV \cite{2020Fujita} and $\cong 10.8$ MeV \cite{2009Yako}.
Several values for the IV-PV coupling constant are used, i.e., $\alpha_{\rm IV-PV}=0$, $1.0$, $1.7$, and $2.1$ fm$^2$.
For the GT($-$) mode, the excitation energy is presented with respect to the ground state of the daughter $^{48}$Sc nucleus.
(Bottom panel) The same plot but for $^{208}$Pb.
The experimental M1 and GT($-$) energies are measured as $E_{\rm M1} \cong 7.3$ MeV \cite{1988Lasz_M1_Exp} and $E_{\rm GT}=15.6$ MeV \cite{1994Akimune, 2012Wakasa}.
} \label{fig:2347}
\end{center} \end{figure}


\subsection{IV-PV residual interaction}
Next we employ the IV-PV Lagrangian density $\mathcal{L}_{\rm IV-PV}$, which contributes at the level of the R(Q)RPA residual interaction \cite{2005Vret, 2005Niksic, 1996Podo}.
In Fig. \ref{fig:2347}, the M1 and GT distributions are presented for $^{48}$Ca and $^{208}$Pb, by using the DD-PC1 and DD-PCX interactions combined with 
the IV-PV interaction term with the coupling constant, $\alpha_{\rm IV-PV}$.
Notice that the pairing correlations vanish in these closed-shell systems.

As a general result in both GT($-$) and M1 cases in Fig. \ref{fig:2347}, the IV-PV interaction works as an additional repulsion, which increases the $1^+$-excitation energies. 
Namely, the single-particle energies and their SO-splitting gaps depend on the spin-parity value of the total system, $J^{P}$, because the IV-PV interaction becomes active (inactive) for the $1^+$ ($0^+$) states \cite{2005Vret, 2005Niksic, 1996Podo}.

For the GT($-$) transition of $^{48}$Ca $\longrightarrow$ $^{48}$Sc, the recent experiment by Fujita et. al. reports the first major peak at $E_{\rm GT}=2.517$ MeV in the daughter $^{48}$Sc nucleus as the low-lying GT excitation \cite{2020Fujita}, whereas the second, giant GT resonance is found around $9-14$ MeV with wide fragmentation \cite{2020Fujita, 2009Yako}.
Compared with this set of experimental data, our calculation becomes consistent when 
the IV-PV coupling, $\alpha_{\rm IV-PV}=1.70$ fm$^2$, is used with the DD-PC1 interaction: the RRPA results in $E_{\rm x}=2.53$ and $13.43$ MeV for the two GT($-$) peaks. 
Note that the second GT($-$) peak is expected as the transition to the continuum above the proton-separation threshold, $9.45$ MeV, in $^{48}$Sc \cite{NNDCHP}.
Therefore, as measured in Refs. \cite{2009Yako, 2020Fujita}, its width can be more fragmented than our smearing assumption.
For elucidating this fragmentation, one may need to employ e.g. the continuum effects \cite{2009Mizuyama,2012Mizuyama} and/or the higher-order configurations \cite{1991Migli,2008Marcucci,1994Moraghe,1982Bertsch,2006Ichimura,2020Gamba}, which go beyond the present RQRPA approach.
On the other side, with respect to the experimental M1-excitation energy, $E_{\rm M1}=10.23$ MeV in $^{48}$Ca \cite{2011Tompkins}, the present setting of $\alpha_{\rm IV-PV}=1.70$ fm$^2$ is a good approximation, i.e., the RRPA with DD-PC1 gives the M1-excitation energy of $9.42$ MeV.
Considering the general accuracy of IV-PV parameter, however, we should mention that the simultaneous reproduction of M1 and GT energies for various nuclei is still demanding.
To further exemplify this point, the results for GT and M1 strength functions for $^{208}$Pb are presented in the lower panel of Fig. \ref{fig:2347}.
One can observe that the current setting of $\alpha_{\rm IV-PV}=1.70$ fm$^2$ with DD-PC1 overshoots its M1 and GT($-$) energies by $1-2$ MeV.

\begin{figure}[t] \begin{center}
  \includegraphics[width = \hsize]{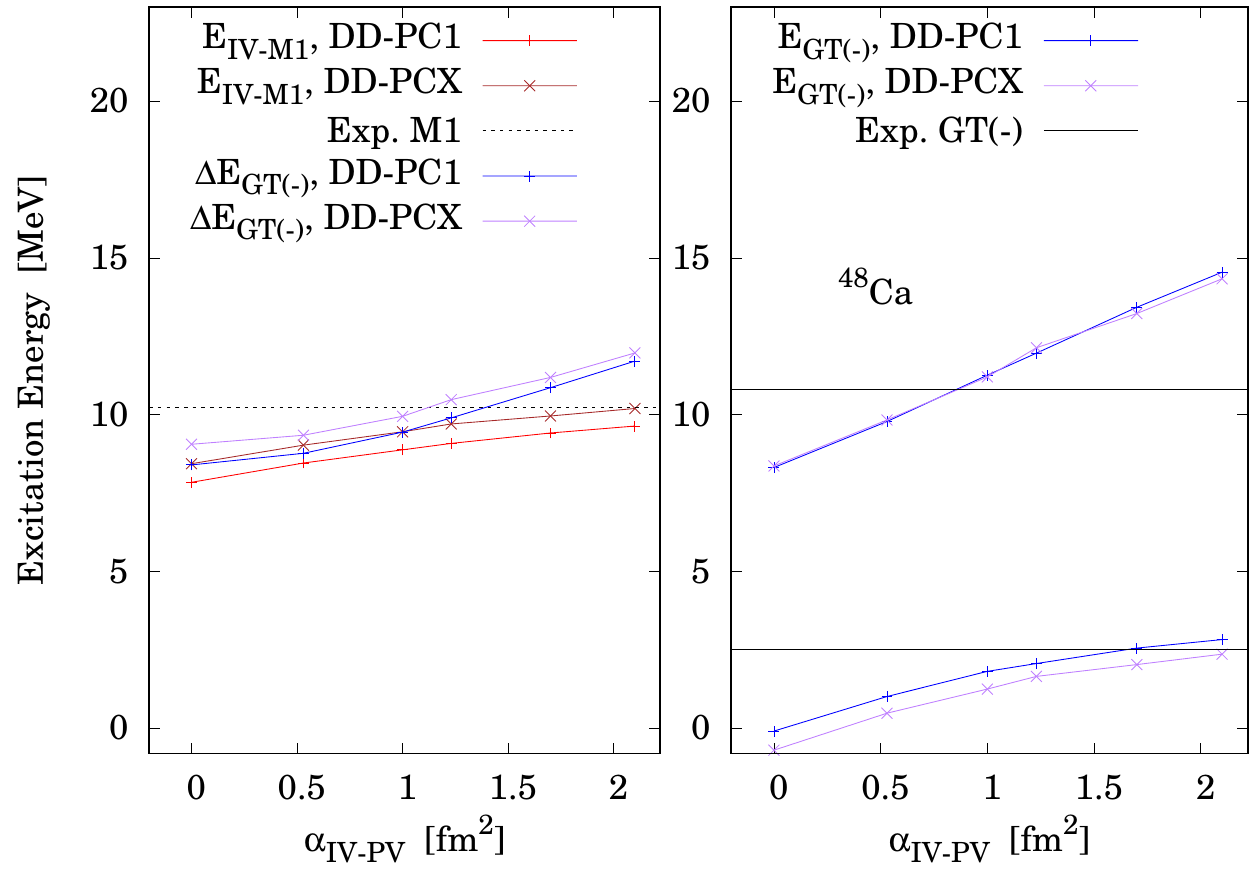}
  \caption{
The M1 and GT($-$) excitation energies for $^{48}$Ca obtained with the DD-PC1 and DD-PCX interactions supplemented with the IV-PV residual interaction for the range of values of its coupling constant $\alpha_{\rm IV-PV}$.
The $\Delta E_{\rm GT(-)}$ denotes the gap between two major GT($-$) energies.
The experimental M1 energy, $10.23$ MeV \cite{2011Tompkins}, is indicated by a dashed line.
The experimental low-lying and giant GT($-$) energies, $2.517$ MeV \cite{2020Fujita} and $10.8$ MeV \cite{2009Yako}, respectively, are indicated by solid lines.
} \label{fig:2348}
\end{center} \end{figure}

The similar problem occurs when we employ the DD-PCX interaction \cite{2019Yuksel} as shown in Fig. \ref{fig:2347}.
For the $^{48}$Ca nucleus, by using $\alpha_{\rm IV-PV}=2.1$ fm$^2$ for the IV-PV coupling, the experimental M1 and low-lying GT energies are well reproduced. 
Simultaneously, however, this result overshoots the experimental energy of giant GT resonance, $E_{\rm GT} \cong 10.8$ MeV in $^{48}$Sc \cite{2009Yako}.
Also, for $^{208}$Pb as another example, the present setting does not match with the experimental M1 and GT energies \cite{1988Lasz_M1_Exp,2012Wakasa}.
Consequently, the simultaneous reproduction of M1 and GT energies for light and heavy systems is challenging in the present context of DD-PCX plus IV-PV interaction.
For quantitative agreement, one may need e.g. the system-dependent tuning of $\alpha_{\rm IV-PV} \longrightarrow \alpha_{\rm IV-PV}(N,Z)$ and/or more complicated parameterization of the IV-PV Lagrangian. 
These tasks go beyond our present scope.
Note also that the present IV-PV coupling constant adjusted to the GT energy in $^{48}$Ca is larger than that in our previous works \cite{2020Oishi,2020Kruzic}.
This is because, in Refs. \cite{2020Oishi,2020Kruzic}, the parameter was adjusted differently, i.e., to the mean value of M1 energies of $^{48}$Ca and $^{208}$Pb nuclei.

Figure \ref{fig:2347} describes how the IV-PV interaction disrupts the equivalence of strength between the M1 and the giant, higher-energy GT($-$) transitions.
Remember that, in the non-perturbed case with $\alpha_{\rm IV-PV}=0$, these two peaks can be attributed to the transition between $(1f_{7/2})$ and $(1f_{5/2})$ states to yield the equivalent $B_{\rm M1/GT}$ values.
Then, with $\alpha_{\rm IV-PV} \ne 0$, the M1-strength height $B_{\rm M1}$ decreases when the IV-PV coupling becomes stronger \cite{1991Migli}, whereas in contrast the higher-energy GT($-$) strength $B_{\rm GT}$ increases \cite{2004Ma}.
This opposite behavior can be attributed to the fact that the IV-PV residual interaction in the non-relativistic limit yields the spin-isospin form of $(\vec{s}_1 \cdot \vec{s}_2) (\vec{t}_1 \cdot \vec{t}_2)$ \cite{2004Ma,2020Oishi}, and thus, 
enhances the isospin-singlet, $\nu(1f_{7/2}) \longrightarrow \pi(1f_{5/2})$ component in the GT($-$) case.
In parallel, the $\nu(1f_{7/2}) \longrightarrow \nu(1f_{5/2})$ component in the M1 excitation is suppressed because the PV interaction does not support this isospin-triplet component. 
Notice also that the low-lying GT($-$) strength is suppressed, namely, the PV interaction does not promote the $\nu(1f_{7/2}) \longrightarrow \pi(1f_{7/2})$ component.
A similar change of GT strength with the residual interaction has been reported in Refs. \cite{2004Ma, 2009Bai}.

Figure \ref{fig:2348} displays the M1 and GT-excitation energies for $^{48}$Ca, calculated with the RRPA using DD-PC1 and DD-PCX interactions, and the IV-PV residual interaction for the range of values of its strength parameter $\alpha_{\rm IV-PV}=0-2.1$ fm$^2$.
If the simple shell-model picture is a good approximation, the M1-excitation energy, $E_{\rm M1}$, is consistent to the SO splitting of neutron levels \cite{2010Heyde_M1_Rev, 2008Pietralla_Rev}, whereas the gap of giant and low-lying GT energies, $\Delta E_{\rm GT(-)}$, corresponds to the proton SO splitting \cite{2007Fujita_GT_Exp,2011Fujita_M1_GT}.
As confirmed in the previous non-perturbed results with $\alpha_{\rm IV-PV}=0$, the $E_{\rm M1}$ and $\Delta E_{\rm GT(-)}$ values coincide except a small gap due to the Coulomb-repulsion potential on the proton side.
We checked that these values are indeed equivalent to the SO-splitting gaps of neutrons and protons in the numerical ground-state solutions.
Then, by increasing the $\alpha_{\rm IV-PV}$ value, the difference of $E_{\rm M1}$ and $\Delta E_{\rm GT(-)}$ values becomes wider.
This tendency is confirmed both in the DD-PC1 and DD-PCX cases.
Consequently, the IV-PV residual interaction affects both M1 and GT energies but in different ways.
This inequality is attributed to the spin-isospin character of the IV-PV interaction, whose effect is not common between the NN and PN configurations in the M1 and GT transitions, respectively.

\begin{figure}[t] \begin{center}
  \includegraphics[width = \hsize]{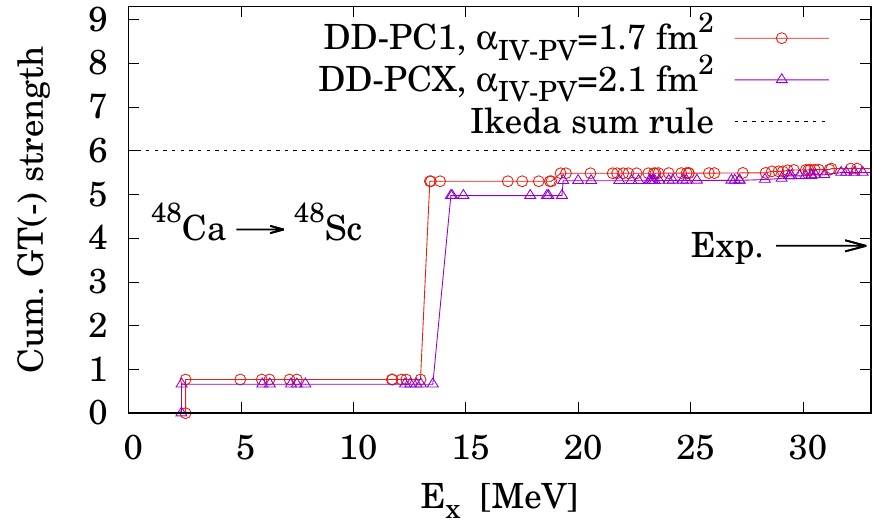}
  \caption{Cumulative GT($-$) strength for $^{48}$Ca obtained from our RRPA method.
Ikeda-Fujii-Fujita sum rule as in Eq. (\ref{eq:IFFSR}) is plotted by the dotted line.
The arrow indicates the experimental value of $\sum B_{\rm GT(-)} = 15.3/4$ up to $30$ MeV \cite{2009Yako}.
} \label{fig:2955}
\end{center} \end{figure}

In Fig. \ref{fig:2955}, the cumulative GT($-$) strength for $^{48}$Ca is plotted.
This is defined as \cite{2014Bai,2020Gamba},
\beqa
 S_{\rm GT(-)}(E_{\rm x}) &=& \int_{0}^{E_{\rm x}} dE \frac{dB_{\rm GT(-)}(E)}{dE} \nonumber  \\
 &=& \int_{0}^{E_{\rm x}} dE R_{\rm GT(-)}(E),
\eeqa
where $R_{\rm GT(-)}(E)$ is the response function in Eq. (\ref{eq:response}).
Also, for comparison, it is convenient to use the Ikeda-Fujii-Fujita sum rule \cite{1963Ikeda, 1964Fujita}.
That is,
\beq
 \lim_{E_{\rm x} \rightarrow \infty} \left[ S_{\rm GT(-)}(E_{\rm x}) -S_{\rm GT(+)}(E_{\rm x}) \right] = \frac{3(N-Z)}{4}, \label{eq:IFFSR}
\eeq
from the total GT($\pm$) summations.
Notice the factor $1/4$ because we use the spin operator $\hat{s}$ instead of $\hat{\sigma}=2\hat{s}$.
For the Ca nuclei, the GT($+$) response is strongly suppressed due to the Pauli blocking effect, namely, $S_{\rm GT(+)}(E_{\rm x} \rightarrow \infty)\cong 0$.
In Fig. \ref{fig:2955},
our cumulative GT strength is consistent with the Ikeda-Fujii-Fujita sum rule, $S_{\rm GT(-)}(E_{\rm x} \rightarrow \infty)\cong 6$ for $^{48}$Ca, with a small deficiency of $\cong 8$\% due to omitting the Dirac-sea states \cite{2004Paar}.
There are two main jumps at $E_{\rm x} \cong 2.5$ MeV and $E_{\rm x} \cong 13$ MeV in correspondence with the GT-excitation energies in Fig. \ref{fig:2347}.
Note that the similar result is obtained from the non-relativistic RPA \cite{2020Gamba}.
However, the experimental data yields $\sum B_{\rm GT(-)} = 15.3 \pm 2.2$ up to $30$ MeV \cite{2009Yako}, which is lower than our result, $4S_{\rm GT(-)}(E_{\rm x} \le 30~{\rm MeV}) \cong 22$.
This discrepancy is due to the missing high-energy GT strength in experiments, and thus, the Ikeda-Fujii-Fujita sum rule is not shown as applicable.
Indeed in Ref. \cite{2020Gamba}, by employing the subtracted second RPA, the GT strength is shown to be more fragmented and in better agreement with experimental data \cite{2009Yako}.
The similar method in the REDF framework, however, is technically challenging, and beyond the present study.

\begin{figure}[t] \begin{center}
  \includegraphics[width = \hsize]{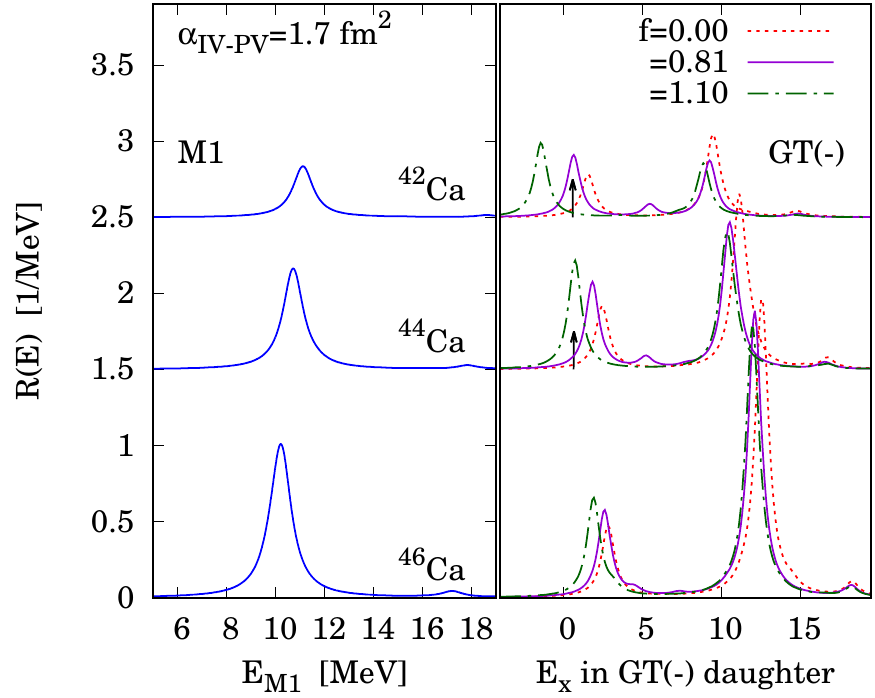}
  \caption{The IV-spin-M1 and GT($-$) strength distributions for $^{42-46}$Ca nuclei with the DD-PC1 and the IV-PV interaction with $\alpha_{\rm IV-PV}=$1.7 fm$^2$.
The GT($-$) response is shown for three values of PN pairing strength, $f=0$, $0.81$, and $1.1$.
Vertical arrows indicate the low-lying GT($-$) energies in experiment \cite{2014Fujita,2020Fujita}: for $^{42}$Ca $\longrightarrow ^{42}$Sc, $0.611$ MeV; for $^{44}$Ca $\longrightarrow ^{44}$Sc, $0.667$ MeV.
} \label{fig:7728}
\end{center} \end{figure}

\subsection{Open-shell Ca isotopes} \label{sec:openshell}
In the following sections, except when modified, we commonly use the DD-PC1 plus IV-PV Lagrangian with $\alpha_{\rm IV-PV}=1.70$ fm$^2$ adjusted to the experimental M1 and GT energies of $^{48}$Ca.
For open-shell nuclei, the pairing correlations should be taken into account.
As explained in Sec. \ref{Sec:Form}, the T1 pairing is described by the pairing part of the Gogny-D1S force as used in Refs. \cite{2014Niksic,2020Kruzic}.
The PN (T0) pairing is described by the two-Gaussian potential as given in Refs. \cite{2020Yuksel,2020Ravlic}.

Figure \ref{fig:7728} shows our results for $^{42-46}$Ca.
In Fig. \ref{fig:7728}, the GT($-$) response is shown for three values of PN pairing strength, $f=$0.0, 0.81, and 1.1 in Eq. (\ref{eq:PNPAIR}).
This PN pairing indeed plays an essential role to reproduce the low-lying GT energy, which appears higher than the experimental data when the PN pairing is neglected.
For instance, the PN pairing with $f=0.81$ reproduces the low-lying GT($-$) excitation of $^{42}$Sc at $0.611$ MeV \cite{2014Fujita,2020Fujita}.
Note that the similar GT distribution and the decrease of GT($-$) energy by the PN-pairing interaction were predicted from the Skyrme QRPA \cite{2014Bai}.
For $^{44}$Ca, however, there still remains a finite gap between the calculated and experimental low-lying GT energies: the enhancement of $f=1.1$ is instead necessary for this system.
It suggests that, instead of the PN-pairing strength commonly used for all the isotopes, one needs the system-dependent fine-tuning of PN-pairing strength and/or further complicated model for quantitative agreement.
This observation agrees with applications of the RQRPA in the calculation of beta decay half-lives, where it was found that a single value of PN pairing strength cannot reproduce experimental half-lives across the isotopic chain \cite{NIU2013_PLB,2016Marketin,2021Ravlic_01,2021Ravlic_02}.

In Fig. \ref{fig:7728}, the absolute strengths of the M1 and giant-GT transitions from $^{42}$Ca become comparable by using the PN-pairing interaction with $f=0.81$:
$R_{\rm M1}=0.34$ MeV$^{-1}$ at $E_{\rm M1}=11.1$ MeV in $^{42}$Ca, whereas
$R_{\rm GT(-)}=0.37$ MeV$^{-1}$ at $E_{\rm x}=9.3$ MeV in the GT($-$) daughter $^{42}$Sc nucleus.
Namely, the M1-GT equivalence, which was disrupted by the IV-PV interaction for closed-shell nuclei in Fig. \ref{fig:2347}, is in this case restored.
On the other hand, that restoring effect is less visible in the neutron-rich Ca nuclei.
In the $^{44}$Ca case, for instance, a stronger attractive PN pairing 
with $f=1.1$ is necessary to reproduce its experimental low-lying GT energy. 
However, there is still a finite difference between its M1 and giant-GT strengths: 
$R_{\rm M1}=0.66$ MeV$^{-1}$ at $E_{\rm M1}=10.72$ MeV in $^{44}$Ca, whereas
$R_{\rm GT(-)}=0.91$ MeV$^{-1}$ at $E_{\rm x}=10.35$ MeV in $^{44}$Sc in Fig. \ref{fig:7728}.

Notice that, by using $f=1.1$, the low-lying GT($-$) response in $^{42}$Sc shows the negative excitation energy. 
Namely, this GT($-$) state results lower than the experimental ground state, because of the over enhancement of the PN-pairing attractive interaction.


\begin{figure}[t] \begin{center}
  \includegraphics[width = \hsize]{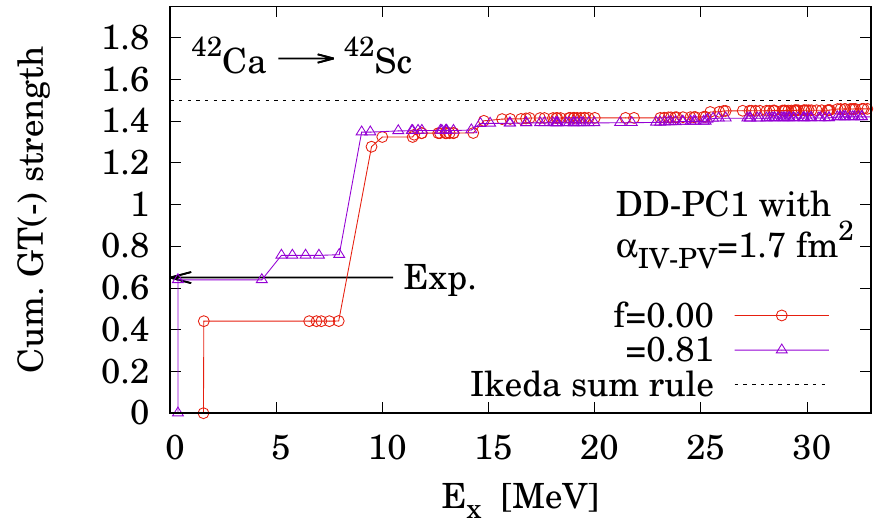}
  \caption{Cumulative GT($-$) strength for $^{42}$Ca obtained with the DD-PC1 plus IV-PV interaction with $\alpha_{\rm IV-PV}=1.7$ fm$^2$.
Ikeda-Fujii-Fujita sum rule as in Eq. (\ref{eq:IFFSR}) is plotted by the dotted line.
The arrow indicates the experimental value of $\sum B_{\rm GT(-)}=2.6/4$ \cite{2014Fujita}.
} \label{fig:6621}
\end{center} \end{figure}

In Fig. \ref{fig:6621}, the cumulative GT($-$) strength for $^{42}$Ca is plotted.
Our cumulative GT strength is consistent with the Ikeda-Fujii-Fujita sum rule, $S_{\rm GT(-)}(E_{\rm x} \rightarrow \infty)\cong 1.5$ for $^{42}$Ca, whereas the GT($+$) response is negligible due to the Pauli blocking effect.
With $f=0.81$ for the PN pairing, there are two main jumps at $E_{\rm x} \cong 0.6$ and $\cong 9$ MeV in correspondence with the two major peaks shown in Fig. \ref{fig:7728}.
However in recent experimental data \cite{2014Fujita}, the second jump has not been confirmed.
In Ref. \cite{2014Fujita}, the measured value of total GT strength for $^{42}$Ca $\longrightarrow$ $^{42}$Sc is $2.6/4 = 0.65$, where the factor $1/4$ is needed for the present comparison.
This value is close to our result, $0.64$ with $f=0.81$ but only from the first peak at $E_{\rm x}=0.6$ MeV in Fig. \ref{fig:6621}.
Namely, the RQRPA calculation can be valid but mainly in the low-lying region.
One possible reason for the absence of experimental GT($-$) strength especially in the high-energy region is the continuum coupling, where the GT($-$) excitation brings the final-state proton above the one-proton threshold, and thus, its width can be large.
Another possibility is the effect of higher-order configurations beyond the present RQRPA level, as we mentioned in the $^{48}$Ca case \cite{2020Gamba}.


\begin{figure}[t] \begin{center}
  \includegraphics[width = \hsize]{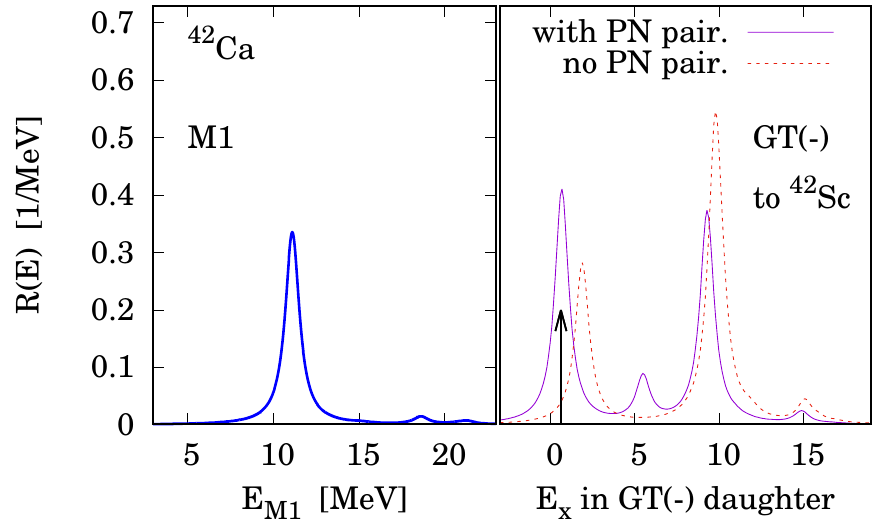}
  \includegraphics[width = \hsize]{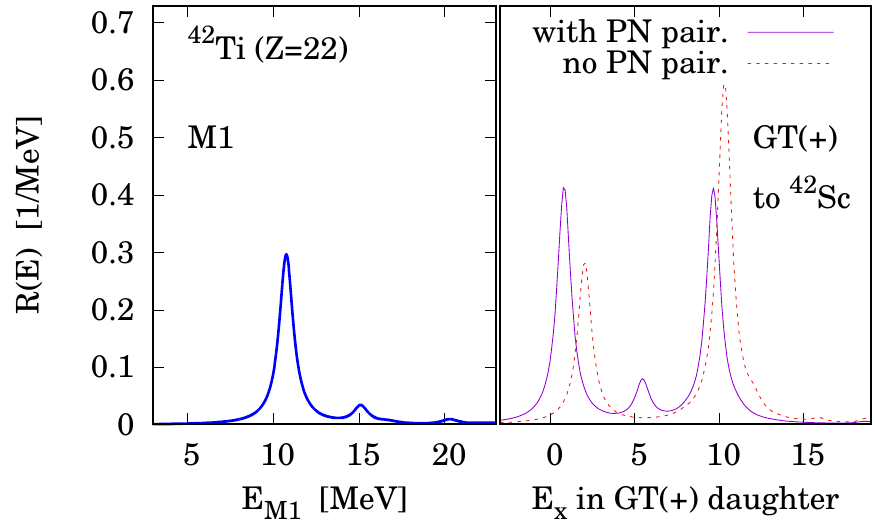}
  \caption{The IV-spin-M1 and GT strength distributions of $^{42}$Ca and $^{42}$Ti obtained with the DD-PC1 plus IV-PV interaction with $\alpha_{\rm IV-PV}=1.7$ fm$^2$.
For GT($\pm$) modes, their excitation energies are presented with respect to the common daughter nucleus $^{42}$Sc. 
The arrow indicates the experimental low-lying GT($-$) energy $0.611$ MeV in $^{42}$Sc \cite{2020Fujita}. } \label{fig:5383}
\end{center} \end{figure}

Before closing our discussion, we check the mirror symmetry in terms of the M1 and GT transitions. 
For this purpose, calculations are performed for ${}^{42}$Ti, which is the mirror system of ${}^{42}$Ca.
Figure \ref{fig:5383} shows our results. 
The GT$(+)$ strength function from $^{42}$Ti coincides well with the GT$(-)$ strength function from $^{42}$Ca, where their daughter nucleus is common, $^{42}$Sc.
Especially the RQRPA predicts the low-lying GT($+$) state equivalently to the GT($-$) case. 
The PN pairing is again found as essential to reproduce its energy. 
For M1 transition, these two isobaric-analog nuclei have consistent M1-excitation energy and strength. 
This can be naturally understood from that their M1 transitions are mainly attributed to the $f_{7/2} \longrightarrow f_{5/2}$ transition in the proton (neutron) side for $^{42}$Ti ($^{42}$Ca).
Consequently, the mirror symmetry holds both in the M1 and GT($\pm$) transitions.
However, we note that the M1-excitation energy in $^{42}$Ti predicted here is much higher than the one-proton-separation threshold, $3.75$ MeV \cite{NNDCHP}, and thus, its fragmentation can be wider than the width used in smearing our RQRPA spectra.
Measurement of this continuum M1 state is thus expected to be challenging.

\section{Summary}
We have discussed the isobaric-analog symmetry between the M1 and GT transitions in Ca isotopes by investigating their excitation energies and transition strengths.
The symmetry of unperturbed M1 and GT response can be broken by 
including the IV-PV residual interaction in the R(Q)RPA, which affects 
the M1 and GT strengths inequivalently.
This is explained by the spin-isospin character of the interaction, whose effect is 
not common for the NN and PN configurations in the M1 and GT transitions, respectively.
The isoscalar PN-pairing effect is also considered for open-shell nuclei.
This effect is found indispensable to reproduce their low-lying GT energies \cite{2014Bai,NIU2013_PLB,2016Marketin}.
Due to the interplay between the IV-PV and pairing channels in the RQRPA, the GT-M1 symmetry can be restored in the open-shell $^{42}$Ca nucleus, i.e., the giant GT excitation shows its strength being comparable to that of the M1 excitation.
A remarkable agreement between the GT($\pm$) transitions is predicted for $^{42}$Ca and $^{42}$Ti nuclei, implying the mirror symmetry.
This mirror symmetry is predicted also in their M1 transitions.

The validity of IV-PV parameterization is also discussed with respect to the experimental M1 and GT-excitation energies.
We conclude that, within the present DD-PC1 (DD-PCX) plus IV-PV parameterization, the accurate and simultaneous reproduction of M1 and GT-transition energies for various nuclei is still rather challenging.
This problem suggests that further improvements of the REDF, especially for the SO splittings, are required, as well as more advanced formulation of the PV coupling.

The quenching effect is not considered in this work.
Whether the common quenching can be valid or not for IV-M1 and GT modes is an open and essential problem, to which we are aiming to approach.
The similar question has been discussed between the IS and IV modes of M1 \cite{2015Matsubara}, for which the RQRPA could answer in near future.
We also note that it could also be interesting to discuss the GT/M1 properties in deformed nuclei. 
However, this task requires a development of the deformed relativistic QRPA and intensive computations.
We aim to address this topic in our future work.

\begin{acknowledgements}
We especially thank Yoshitaka Fujita, Hirohiko Fujita, Kosuke Nomura, and Esra Y\"uksel for fruitful discussions.
\modific{4}{T.O. acknowledges the support of the Yukawa Research Fellow Programme by Yukawa Memorial Foundation.}
This work is financed within the Tenure Track Pilot Programme of the Croatian Science Foundation and the \'{E}cole Polytechnique F\'{e}d\'{e}rale de Lausanne, and the Project TTP-2018-07-3554 Exotic Nuclear Structure and Dynamics, with funds of the Croatian-Swiss Research Programme.
This work is supported by the ``QuantiXLie Centre of Excellence'' project co-financed by the Croatian Government and European Union through the European Regional Development Fund, the Competitiveness and Cohesion Operational Programme (code KK.01.1.1.01.0004).
Numerical calculations were supported by the Multidisciplinary Cooperative Research Program of the Center for Computational Sciences, University of Tsukuba, using Oakforest-PACS Systems (project No. xg21i064, FY2021), and T.O. acknowledges the support by Takashi Nakatsukasa and Hiroyuki Kobayashi in this program.
\end{acknowledgements}

\appendix
\section{Inclusion of orbital M1 component}
In this section, we investigate the effect of orbital M1 component.
The original IV-M1 operator is given in Eq. (\ref{eq:SPWUE}) in the main text, 
including both $\hat{\tau}_0 \hat{s}_{\nu}$ and $\hat{\tau}_0 \hat{l}_{\nu}$ terms.
We perform the RQRPA calculations by using these operators.
Note that the factor $\mu_{\rm N} \sqrt{3/4\pi}$ and g factors, $g_s=4.706$ and $g_l=1/2$, are taken into account in this section.
The DD-PC1 and T1-pairing interactions are utilized samely to the main text.
The IV-PV coupling of $\alpha_{\rm IV-PV}=1.7$ fm$^2$ is also used in the R(Q)RPA-residual interactions.

\begin{figure}[t] \begin{center}
\includegraphics[width = \hsize]{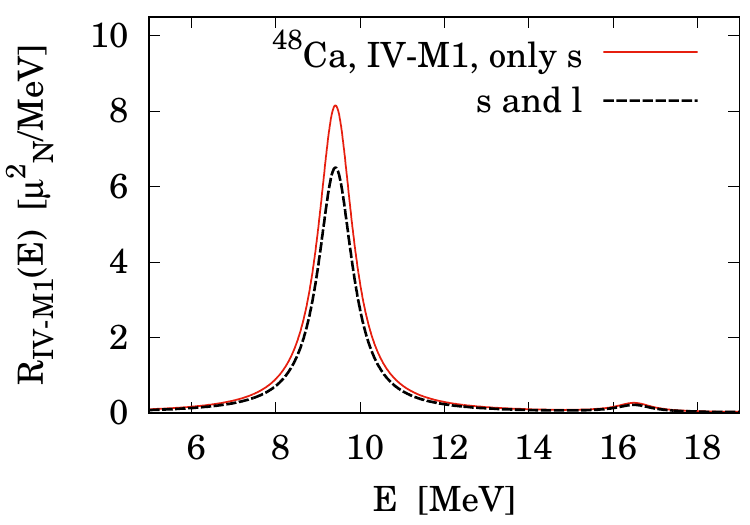}
\includegraphics[width = \hsize]{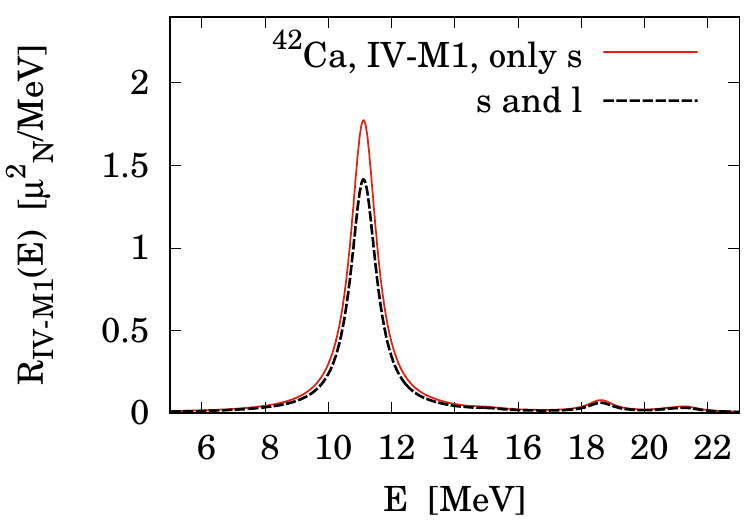}
\caption{The IV-M1 response functions for $^{42,48}$Ca nuclei.
} \label{fig:3633}
\end{center} \end{figure}

In Fig. \ref{fig:3633}, our results for the $^{48}$Ca and $^{42}$Ca nuclei are displayed.
There we compare the two cases, where the orbital-M1 operator, 
$g_l \mu_{\rm N} \sqrt{3/4\pi} \cdot \hat{\tau}_0 \hat{l}_{\nu}$ is neglected (only s) or included (s and l).
It is shown that the inclusion of orbital-M1 component reduces the IV-M1 strength of these nuclei.
We checked that, in terms of M1-response summation, which reads $\int R_{\rm M1}(E) dE$, this reduction is approximately $80$\% of the ``only s'' case.
Since the orbit-M1 inclusion is shown as another source of quenching of $B_{\rm M1}$, a careful treatment is necessary in forthcoming studies.
On the other hand, the major excitation energies are not affected even if the orbital-M1 component is included.
This is natural because the selection rule of spin and orbital IV-M1 operators are identical \cite{2007Suhonen}.
Therefore, the active excited states in the R(Q)RPA calculations should be common and have the same eigenenergies.

%


\end{document}